\documentclass[11pt]{article}
\usepackage{geometry}                
\usepackage{graphicx}
\usepackage{amssymb}
\usepackage{epstopdf}
\usepackage[auth-sc]{authblk}
\newcommand{\be}{\begin{eqnarray}}
\newcommand{\ee}{\end{eqnarray}}
\title{Information-theoretic characterization of the complete genotype-phenotype map of a complex pre-biotic world\\ \mbox{} \\
{\small Comment on ``From genotypes to organisms: State-of-the-art and perspectives of a cornerstone in evolutionary dynamics" by Susanna Manrubia et al.}}

\author[1,2]{Nitash C G}
\author[2,3,4,5,*]{Christoph Adami}
\affil[1]{Department of Computer Science and Engineering, Michigan State University}
\affil[2]{BEACON Center for the Study of Evolution Action, Michigan State University}
\affil[3]{Department of Microbiology and Molecular Genetics, Michigan State University}
\affil[4]{Department of Physics and Astronomy, Michigan State University}
\affil[5]{Program in Ecology, Evolution, and Behavior, Michigan State University}

\affil[*]{adami@msu.edu}

\begin{document}
\maketitle
\begin{abstract} How information is encoded in bio-molecular sequences is difficult to quantify since such an analysis usually requires sampling an exponentially large genetic space. Here we show 
how information theory reveals both robust and compressed encodings in the largest complete genotype-phenotype map (over 5 trillion sequences) obtained to date.
\end{abstract}

In the target article ``From genotypes to organisms: State-of-the-art and perspectives of a cornerstone in evolutionary dynamics"~\cite{Manrubiaetal2021} the authors share a broad overview of computational and theoretical investigations into the nature and properties of complex fitness landscapes. These fitness landscapes are, as the authors emphasize throughout, complex maps from a genotype space to a phenotype space. The map is necessarily complex because for the most part, there are many more distinct genotypes (in a sense, exponentially more) than there are distinct phenotypes. One consequence of this feature of the genotype-phenotype (GP) map is that the fraction of functional sequences (those that have a particular phenotype) is very small compared to the number of possible sequences. In fact, this ``density" of functional sequences can be used to quantify the {\em information content} of the sequences, as pointed out by Szostak~\cite{Szostak2003}. He defined the ``functional information" a molecule has about a particular phenotype $E$ in terms of the fraction of sequences $F(E)$ that carry that phenotype at a level $\theta$ or higher, as
\be
I(E_\theta)=-\log F(E\geq\theta)\;, \label{fun}
\ee
where $F(E\geq\theta)=N_\theta/N$, with $N_\theta$ the number of sequences with at least functionality $\theta$, and $N$ is the total number of possible sequences under investigation (typically, constrained by length). If the sequences are written in an alphabet of size $D$ ($D=20$ for proteins, for example) and limited to sequence length $L$, then $N=D^L$ and
\be
I(E_\theta)=L-\log_D N_\theta\;. \label{info}
\ee
Since $\log N_\theta$ is just the ``microcanonical" (or ``coarse-grained") approximation of the Shannon entropy and $L=\log_D(N)$ is the unconditional (maximal) entropy, it turns out that (\ref{fun}) is simply an approximation of the Shannon information content of a biomolecular sequence, introduced earlier in~\cite{AdamiCerf2000} and reviewed in~\cite{Adami2004}. Thus, Szostak's $I(E_\theta)$ is a convenient way 
to quantify how much information about performing task $E$ is stored within a sequence. 

A drawback of $I(E_\theta)$ is that it requires knowledge of the complete genotype-phenotype map to accurately measure the information content. However, approximations to $I(E_\theta)$ exist that allow you to estimate the information content by using single- and two-point substitutions, as done for example in an estimate of the information content of the HIV protease~\cite{GuptaAdami2016}. Here we present an information-theoretic analysis of information encoding in the complete GP map of a ``pre-biotic" world: the sequence space of self-replicators in the digital life system Avida~\cite{AdamiBrown1994,Adami1998,Ofriaetal2009}. In Avida, self-replicating programs can be written in a custom programming language that typically uses 26 instructions, conveniently labeled by the lowercase letters a-z, that can then evolve under natural selection and adapt to complex landscapes. It is known that the smallest self-replicator that can be written in this language requires 8 instructions, and an analysis of all $26^8\approx 209\times 10^9$ sequences revealed exactly 914 sequences that are viable and ``colony-forming", that is, they give rise to a growing population.

\begin{figure}[htbp]    \centering
   \includegraphics[width=3in]{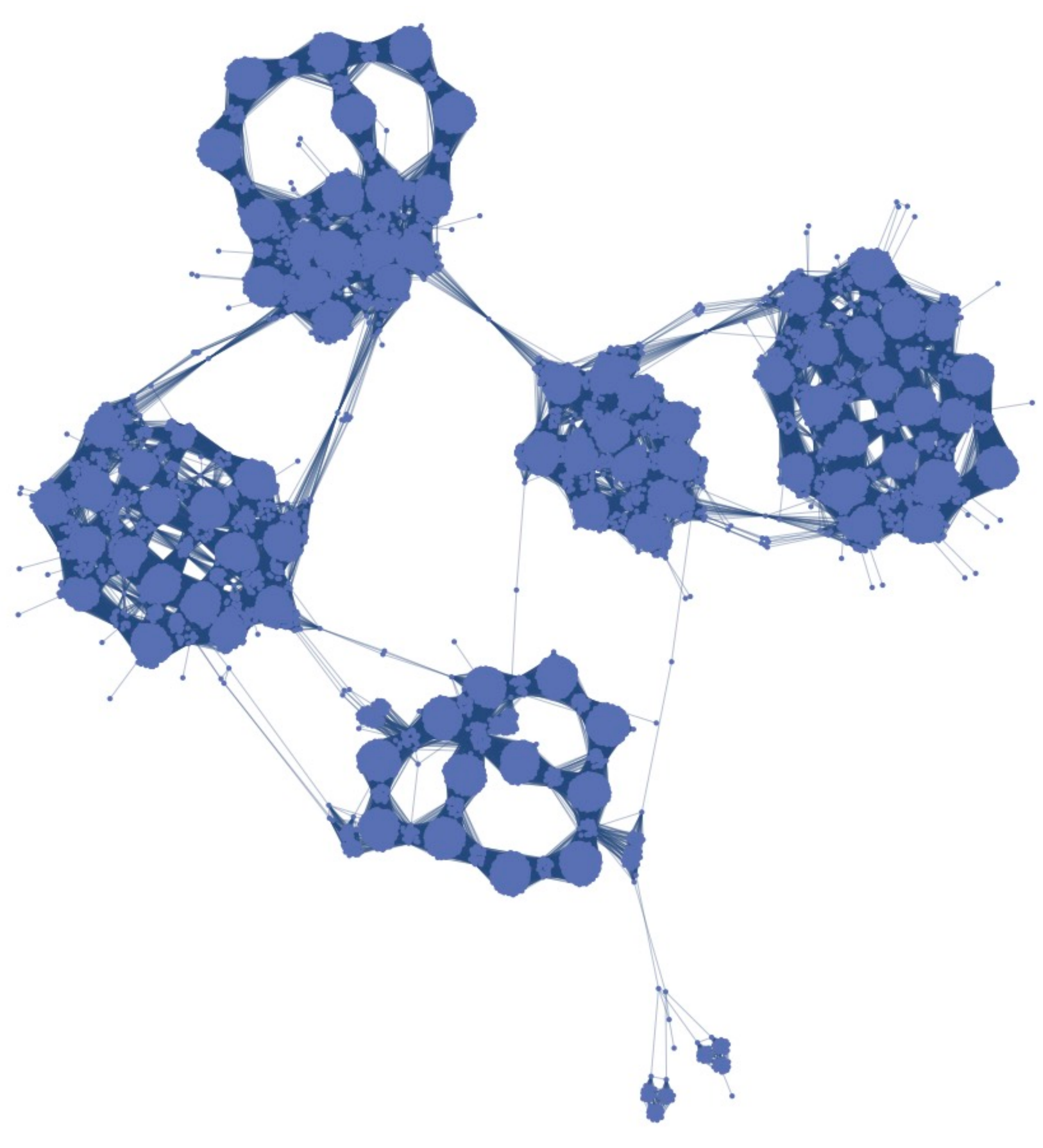} 
   \caption{Largest cluster containing over 94\% of the replicators of the $L=9$ landscape, with edges drawn between one-mutant neighbors. The graph consists of five sparsely connected groups of tightly-connected clusters.  }
   \label{fig1}
\end{figure}

Using this number for $N_\theta$ in (\ref{info}) and using the phenotype  ``colony-forming" as the threshold reveals an information content\footnote{A `mer' is a unit of entropy or information of a polymer in terms of the monomer entropy, by taking logarithms to the base of the alphabet size $D$.}
\be
I_8=-\log_{26}{(914/26^8)}\approx5.91\ {\rm mers}\;.  \label{L8}
\ee
Here we present an analysis of information encoding for the complete landscape of $L=9$ replicators in the Avida landscape. When testing all $26^9=5.43\times 10^{12}$ sequences we found 30,547 replicators, forming 84 distinct connected clusters in sequence space\footnote{There are in fact 36,171 distinct replicators, but some of them are equivalent under a {\em code rotation}. Avidian genomes have a beginning and an end, but the code is circular so two distinct replicators sometimes have the same sequence if one can be rotated into the other. This rotation is reflected in the clusters, but not in the calculation of the information content.}. The (by far) largest of these clusters with 27,753 sequences is shown in Fig.~\ref{fig1}. 
The information content of these replicators is 
\be
I_9=-\log{(36,171/26^9)}\approx 5.77\ {\rm mers}\;. \label{L9}
\ee
Replicators differ vastly in their robustness (as measured by the number of one-mutant neighbors that are also replicators): while the most robust replicator has 74 viable one-mutant neighbors, the least-robust one has none at all (edge distribution not shown). 

\begin{figure}[htbp] 
    \centering
    \includegraphics[width=4.5in]{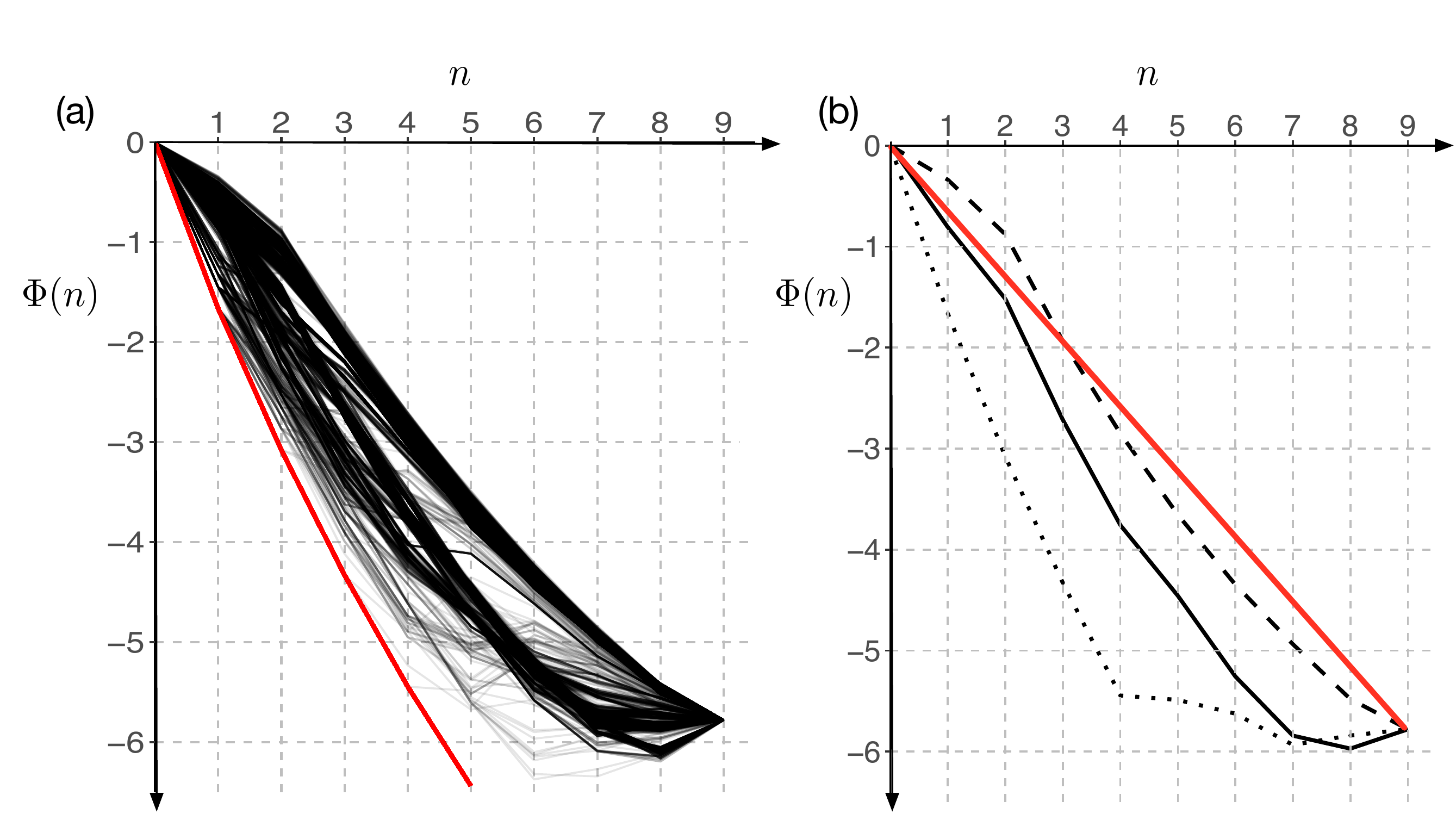} 
    \caption{(a): The log of the fraction of functional sequences, $\Phi(n)$, at mutational depth $n$ for all replicators of $L=9$. The red line shows the theoretical lower limit for a $\Phi(n)$ with perfectly compressed information (no viable mutational neighbors). (b): $\Phi(n)$ for the most fragile (dotted line) and the most robust (dashed line) replicator. The average $\Phi(n)$ across all replicators is indicated by the solid line. The red line indicates the density decay if sites do not interact (no epistasis). }
    \label{fig2}
 \end{figure}

We can test how information is encoded in replicator $i$ by analyzing the density of replicators $n$ mutational steps away from that sequence:
\be
 \rho_i(n)=\frac{\sum_{k=0}^n N^{(i)}_\nu(k)}{\sum_{k=0}^n {L\choose k}25^k }\;.
  \label{nu}
 \ee 
In Eq.~(\ref{nu}), $N^{(i)}_\nu(k)$ is the number of viable replicators at a mutation depth $k$ away from the target sequence $i$ (the denominator $\sum_k {L\choose k}25^k$ is the number of possible mutants up to distance $n$), so that $\rho_i(n)$ is the density of replicators at distance {\em up to} $n$ from sequence $i$. Plotting $\Phi(n)=\log\rho(n)$ against $n$ (Fig.~\ref{fig2}) reveals how information is encoded within the sequence. 
 
A gently decreasing slope of $\Phi(n)$ for small $n$ indicates robust replicators. We can see in Fig.~\ref{fig2} a significant variety of encodings among the replicators, from the most robust (with 74 viable neighbors) to sequences with maximally-compressed encoding up to $n=4$. The most robust replicators show antagonistic (negative) epistasis between the first few sites, while informationally fragile sequences display synergistic (positive) epistasis for the first few mutations, but antagonism for the later ones (see~\cite{WilkeAdami2001}).  All the information-density curves in Fig.~\ref{fig2} must end at the same point, as by definition $\Phi(9)=-I_9$ from Eq.~(\ref{L9}). 

In summary, an information-theoretic analysis of the complete genotype-phenotype map of complex landscapes can reveal a surprising amount of detail about how information is encoded in molecular sequences,  and goes beyond the standard analysis of cluster size and mutational robustness shown in the target article (see also Refs.~\cite{Cowperthwaiteetal2008,SchaperLouis2014,Fortunaetal2017}). However, analyzing larger landscapes will require approximate methods that neglect higher-order epistatic effects, and must rely on a parameterization of the information-density decay function~\cite{WilkeAdami2001}.
\\ \mbox{}\\
\noindent {\bf Acknowledgements} This research was funded by the BEACON Center for the Study of Evolution in Action. We acknowledge computational resources provided by the Institute for Cyber-Enabled Research (iCER) at Michigan State University.

\bibliographystyle{unsrt}
\bibliography{PLR}
\end{document}